\documentclass[aps,floatfix, bibnotes,showpacs,prl,twocolumn]{revtex4}
\usepackage {amsmath}
\usepackage {amssymb}
\usepackage {graphicx}
\usepackage[T2A]{fontenc}
\usepackage{indentfirst}

\usepackage[cp1251]{inputenc}
\usepackage[russian]{babel}

\usepackage[usenames]{color}

\newcommand{\eps}{\varepsilon}
\newcommand{\pa}{\partial}

\renewcommand{\kappa}{\varkappa}

\begin{document}

\title{ Отражение от профилированной границы гиперболического метаматериала}
 \author{Н. А. Жарова}
 \email{zhani@appl.sci-nnov.ru}
 \affiliation{Институт прикладной физики, РАН, Нижний Новгород,  603950 Россия}
 \author{А. А. Жаров}
 \affiliation{Институт физики микроструктур, РАН, Нижний Новгород,  603950 Россия}
 \author{А. А. Жаров, мл. }
 \affiliation{Институт физики микроструктур, РАН, Нижний Новгород,  603950 Россия}

\begin{abstract} 	
Проведено численное исследование особенностей рассеяния падающего излучения на периодически модулированной границе гиперболического метаматериала. Найдена оптимальная - пилообразная -  форма профиля, обеспечивающая минимум отражения. Показано, что при определенной ориентации оптических осей метаматериала на пилообразной границе осуществляется субволновая локализация поля, найдено приближенное аналитическое выражение для собственной моды. Проведено обобщение численного метода расчета полей, основанного на применении  второго тождества Грина.  
\end{abstract}
\maketitle

 \section{Введение}
Большой интерес исследователей к так называемым гиперболическим метаматериалам (ГМ) \cite{h1,h2,h3,h4,Hashemi_PRB_12,Nefedov_SciRep_13,Nefedov_JOpt_13} обусловлен их необычными и полезными для приложений электромагнитными свойствами. 
ГМ это анизотропная резонансная среда, которая характеризуется изочастотной поверхностью гиперболического типа, а не эллипсоидальной, как для обычных сред. В оптике ГМ может быть реализован, например, как металло-диэлектрическая планарная наноструктура, либо как решетка металлических нанопроволок в диэлектрической матрице. 

Уникальные свойства ГМ могут найти множество приложений, среди которых отрицательная рефракция  \cite{23, 24}, создание субволновых изображений  \cite{Jacob_OE_06,Sun_NatCom_15,Tumkur_APL_12,Tumkur1_APL_12,25,Mantel_OL_11},  локализация света в широком частотном диапазоне  \cite{cui_NanoLett12,167}.
 Очень высокая плотность фотонных состояний в ГМ  \cite{Jacob_ApplPhysB_10,Shekhar_NanoConvergence_14,we4} обеспечивает эффективный излучательный теплообмен  \cite{h4,Hashemi_PRB_12,Nefedov_SciRep_13}, что может использоваться в дизайне наноразмерных устройств нагрева/охлаждения.

Гиперболический характер метаматериала достигается лишь при достаточно высокой объемной доле металла в композите, что также означает большие омические потери. 
Поглощение излучения особенно сильно для мелкомасштабных мод, когда фазовая скорость волны направлена вдоль образующей резонансного конуса. Для оптической литографии и создания изображений с субволновым разрешением омические потери нежелательны, но сильное поглощение в ГМ может быть использовано для дизайна поглощающих покрытий. 
В работе \cite{we5} был предложен дизайн поглощающего покрытия на основе неоднородного гиперболического метаматериала. Оказывается, что полное  поглощение в широкой полосе частот можно реализовать в плоском тонком слое  ГМ при условии, что  ориентация оптической оси плавно меняется от параллельной к нормальной (по отношению к интерфейсу). Однако к сожалению в таком тонком слое трудно контролировать направление оптической оси метаматериала (т.е., например, ориентацию металлических проволочных включений). 

Идеальный  поглотитель должен обеспечивать не только 
полное поглощение излучения в возможно более тонком слое покрытия. Очень важно достигнуть минимального отражения на границе с воздухом.  
Известный способ снижения коэффициента отражения это структурирование поверхности поглощающего материала, то есть профилирование его границы с воздухом. 
 В  данной работе мы  изучаем  особенности рассеяния падающего излучения на слое гиперболического метаматериала  с профилированной границей, в частности влияние формы, продольного и поперечного характерного масштаба неоднородности профиля на коэффициент отражения. 

\section{Постановка задачи} 
\begin{figure}[htb]\centering
\includegraphics[width = 0.7\columnwidth]{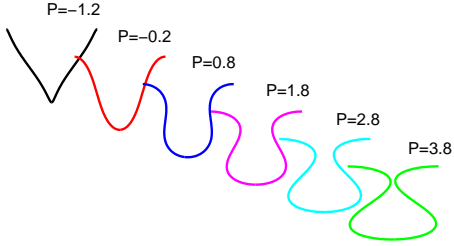}
\caption{\label{fig_curves} (color online) 
Форма профиля $y_B(x_B)$ периодической границы (элементарная ячейка) для различных значений параметра $P$.   
} 
\end{figure}

 В простейшей постановке задача сводится к отражению плоской волны TM поляризации  от двумерного профиля $\hat \eps$. Метаматериал  описывается диэлектрическим тензором $\hat \eps$, и в главных осях $\hat \eps = [\eps_\|, \  0; \  0, \  \eps_\bot ]$. Граница $x_B(\tau ), y_B(\tau )$, разделяющая свободное пространство и метаматериал, является периодической функцией параметра $\tau$ (длина вдоль границы) и характеризуется масштабом вдоль границы $L_x$, амплитудой профиля $L_y$ и формой. 
 Масштабирование профиля по $x$ и $y$ координатам для заданной формы выполняется элементарно, но задача выбора формы не так тривиальна. Оптимальным представляется задание $x_B(\tau ), y_B(\tau )$ математической формулой с параметром, изменение которого существенно меняет форму границы.  
 В дальнейшем мы использовали  функцию, задающую $x_B(\tau ), y_B(\tau )$ на длине элементарной ячейки $\tau = [-\pi /2 : 3/2 \pi]$, в виде 
\begin{align}\label{eq_curve}
 x_B=\pi \int \cos \theta d\tau - P \sin (2\tau ), \quad y_B=\pi \int \sin \theta d\tau ,
 \end{align} 
 где $\theta = -0.4\pi \cos \tau$. 
 Управляющий параметр $P$ меняет форму профиля от почти пилообразной к почти косинусоидальной и до напоминающей меандр. На рисунке \ref{fig_curves} приведены в качестве иллюстрации кривые $y_B(x_B)$ для набора значений $P$.

Перебирая различные значения параметра $P$, масштаба элементарной ячейки вдоль границы $L_x$ и амплитуды профиля $L_y$, можно найти оптимальное сочетание, которое обеспечивает слабое отражение падающего излучения от метаматериала. Дополнительным параметром оптимизации может служить также угол ориентации главных (оптических) осей метаматериала по отношению к осям $x$, $y$. 
        
Коэффициент отражения от полупространства ГМ  с профилированной границей  может быть найден при  решении уравнения Гельмгольца методом конечных элементов (FEM). Однако попытка использовать стандартную FEM программу решения уравнений в частных производных оказалась неудачной. Поскольку в этом случае нет прямого доступа к коду, то возникают проблемы:  появляется  отражение излучения от границ расчетного интервала, приходится  рассматривать отражение не плоской волны, но  широкого гауссова пучка из-за того, что во встроенном модуле нет опции задания периодических по $x$ граничных условий. Это все вносит ошибки в расчет коэффициента отражения.
Более удобным оказался поэтому другой метод расчета,  основанный на использовании второго тождества Грина (Green's second identity), подробное описание которого дается в Приложении.

 \section{Результаты численного исследования}
 Тестирование метода и соответствующей расчетной программы проводилось для предельных случаев (i) слабой модуляции границы, (ii) одинаковых диэлектрических характеристик двух сред, (iii) малого характерного периода модуляции, когда работает приближение эффективной среды. Расчетные и теоретические значения коэффициента отражения совпали в этих случаях с точностью $\sim 1$\% (ошибка  обусловлена дискретизацией вычислений). 
 Кроме того, мы проводили сравнение распределения полей, полученного нашим методом и рассчитанного стандартной программой, которая использует метод конечных элементов. С учетом сказанного выше относительно трудностей применения стандартной программы, это сравнение показало разумное совпадение.

Параметры гиперболической среды, использованные в численном моделировании были найдены для метаматериала представляющего собой диэлектрическую матрицу с металлическими включениями в форме проволок (wire medium).  
В приближении эффективной среды компоненты диэлектрического тензора $\hat \eps$ вдоль проволок $\eps_\| $ и поперек них $\eps_\bot$ находятся как \cite{wire_medium} 
\begin{align}\label{epsilon}
 \eps_\| =\eps_M\rho_M+\eps_D(1-\rho_M), \notag \\
 \eps_\bot=\eps_D \frac{\eps_M (1+\rho_M)+\eps_D (1-\rho_M)}{\eps_D (1+\rho_M)+\eps_M(1-\rho_M)},
\end{align}
где  $\eps_M$ диэлектрическая проницаемость металла, $\eps_D$ проницаемость диэлектрической матрицы и  $\rho_M$ объемная доля металла в составе композита. Далее мы используем для расчетов параметры стекла   ($\eps_D$=2.1590) и серебра с объемной долей   $\rho_M=0.25$. Дисперсионная зависимость проницаемости серебра   $\eps_M (\lambda )$ берется из базы данных  \cite{database}. Для вакуумной длины волны излучения $\lambda = 0.6 \mu$m величина $\eps_M = -15.9822 + 0.5899i$, и расчеты по формулам (\ref{epsilon}) дают значения $\eps_\| =4.2660 + 0.0318i$ и $\eps_\bot = -2.3763 + 0.1475i$.

\begin{figure}[htb]\centering
\includegraphics[width = 0.7\columnwidth]{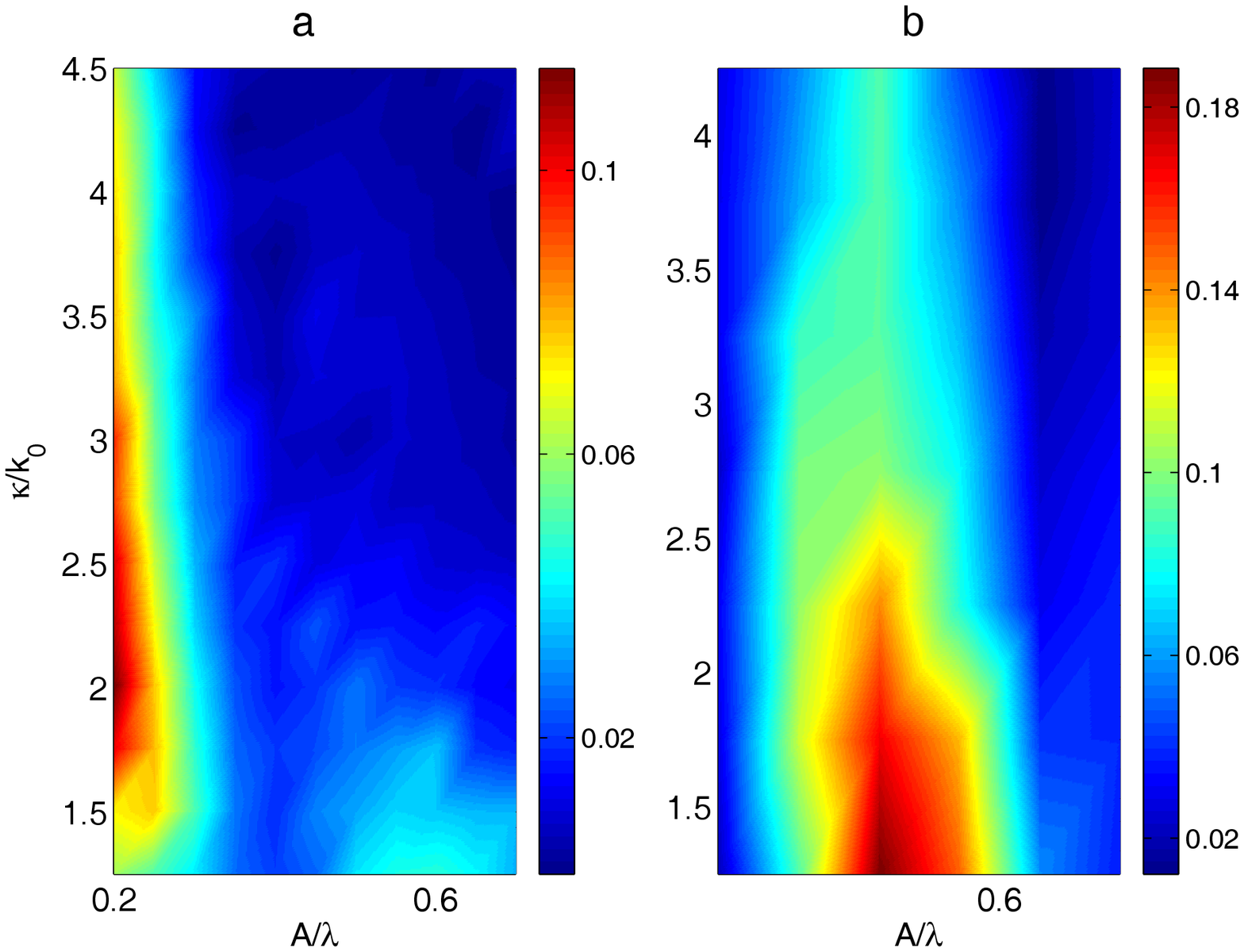}
\caption{\label{fig_kappa_amp0} (color online) 
Коэффициент отражения от профилированной границы с параметром формы $P = -1.2$ (a) и $P = 0.5$ (b) как функция амплитуды модуляции $A/\lambda$ и нормированного волнового числа модуляции $\kappa /k_0 = L_x/\lambda$. }
\end{figure}

Результаты расчета рассеяния излучения при нормальном падении на периодическую границу ГМ с помощью метода, описанного в Приложении, иллюстрируются на рисунке~\ref{fig_kappa_amp0}, где приведены значения  коэффициента отражения как функции амплитуды модуляции $A/\lambda$ и нормированного волнового числа модуляции $\kappa /k_0 = L_x/\lambda$ для двух различных типов профиля, пилообразного (с параметром формы $P = -1.2$) и косинусоидального (с параметром формы $P = 0.5$). 
Если для пилообразного профиля рост амплитуды модуляции приводит к снижению отражения по сравнению со случаем плоской границы, то максимальный коэффициент отражения от косинусоидального профиля превышает его почти в два раза. 
Такое нетривиальное поведение коэффициента отражения обусловлено тензорным и, главным образом, гиперболическим  характером среды.

\begin{figure}[htb]\centering
\includegraphics[width = 0.7\columnwidth]{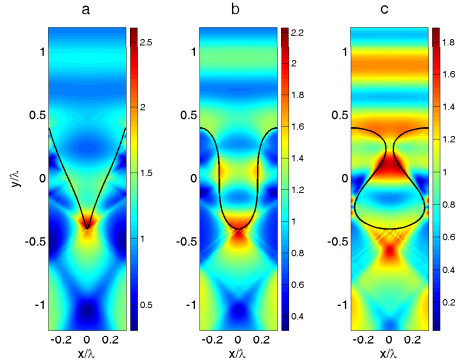}
\caption{\label{fig_fields} (color online) 
Распределение модуля магнитного поля $|h(x,y)|$ (показан один период структры) при нормальном падении плоской волны $h^{inc} = exp(-ik_0y)$ на периодически модулированную границу гиперболического метаматериала. Амплитуда модуляции $A=0.8\lambda$, период структуры в $x$ направлении $\lambda /1.5$, параметр $P$, контролирующий форму модуляции (см. формулу (\ref{eq_curve})) $P = -1.2$, 0.5  и 3.5 соответственно для панелей (a), (b) и (c). Рассчитанный для этих трех вариантов коэффициент отражения $R = $0.016, 0.065 и 0.16. Остальные параметры численного моделирования приведены в тексте. 
} 
\end{figure}

На рисунке~\ref{fig_fields} приведены рассчитанные распределения поля для трех различных типов профиля границы гиперболической среды. Коэффициент отражения плоской волны $h^{inc} = \exp (-ik_0 y)$ оказывается равным $R = $0.016 (панель (a)), 0.065 (панель (b)) и 0.16 (панель (c)). Оптимальной формой границы, обеспечивающей минимум отражения, является пилообразный профиль (см. также рис.~\ref{fig_kappa_amp0}). 
В случае (b) (косинусоидальный профиль) отражение примерно вдвое меньше, чем от плоской границы полупространства с $\eps = \eps_\|$,  но максимальное отражение наблюдается от меандроподобного профиля   (панель(с)), и коэффициент отражения в этом случае заметно больше, чем от плоской границы.

\begin{figure}[htb]\centering
\includegraphics[width = 0.7\columnwidth]{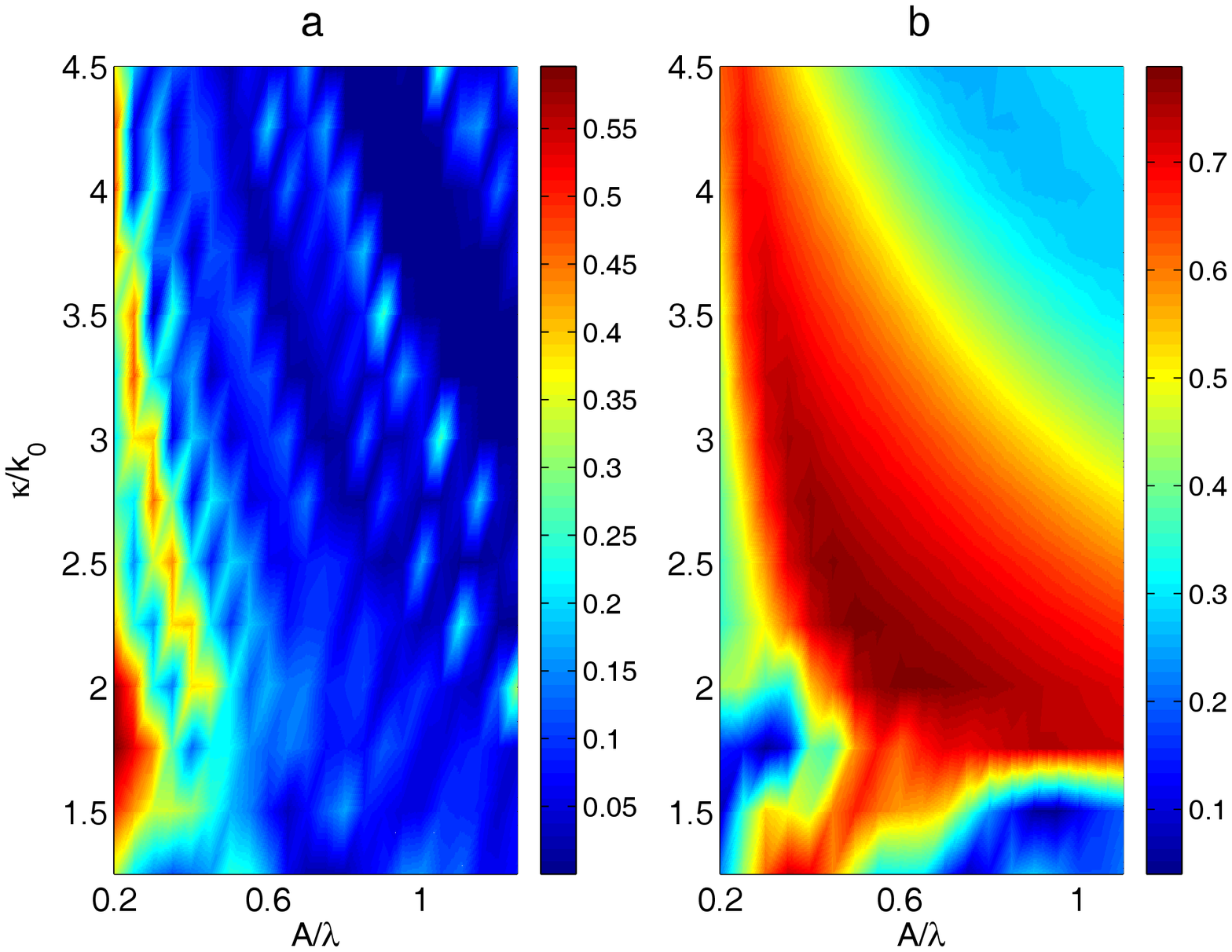}
\caption{\label{fig_kappa_amp} (color online) 
Коэффициент отражения от профилированной границы с параметром формы $P = -1.2$ (a) и $P = 0.5$ (b) как функция амплитуды модуляции $A/\lambda$ и нормированного волнового числа модуляции $\kappa /k_0 = L_x/\lambda$. Ориентация осей анизотропии повернута на 90$^o$ по сравнению со случаем изображенном на рисунке~\ref{fig_kappa_amp0}} 
\end{figure}

Коэффициент отражения существенно зависит от ориентации осей анизотропии метаматериала по отношению к границе. 
При повороте осей на 90$^o$, когда компонента тензора проницаемости вдоль границы отрицательна,  $\eps_{xx} = \eps_\bot = -2.3763 + 0.1475i$, коэффициент отражения от плоской границы  близок к единице, и  профилирование границы уменьшает отражение.  
На рисунке~\ref{fig_kappa_amp}  приведена зависимость коэффициента отражения от профилированной границы с параметром формы $P = -1.2$(a) и $P = 0.5$ (b)  как функция амплитуды модуляции $A/\lambda$ и нормированного волнового числа модуляции $\kappa /k_0 = L_x/\lambda$. 
Существенное уменьшение отражения наблюдается при условии роста как амплитуды модуляции $A$, так и волнового числа модуляции $\kappa$ для $P = -1.2$. Эта же тенденция, хотя и с гораздо меньшим эффектом, прослеживается для профиля другой формы с параметром $P = 0.5$   (форма профиля изображена на рис.~\ref{fig_fields}~(b)).

\begin{figure}[htb]\centering
\includegraphics[width = 0.7\columnwidth]{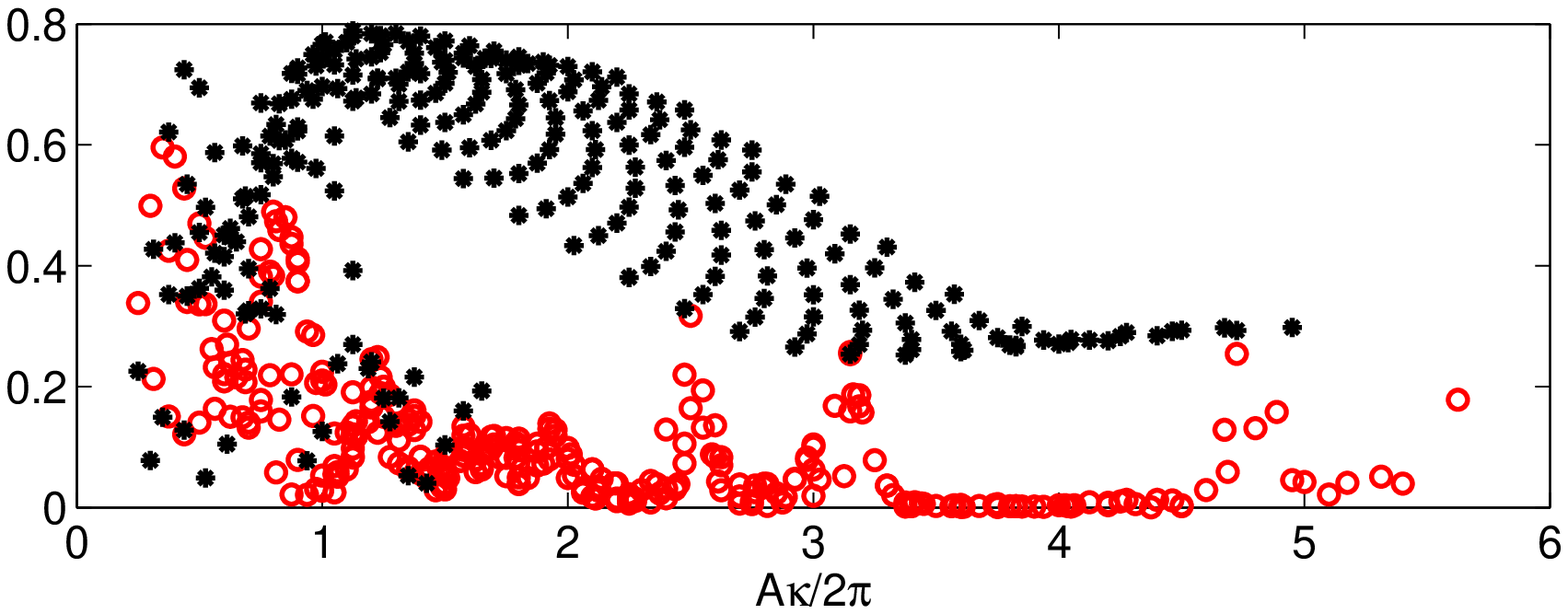}
\caption{\label{fig_kappa_amp1} (color online) 
Коэффициент отражения от профилированной границы с параметром формы $P = -1.2$ (красные кружки) и $P = 0.5$ (черные звездочки) как функция параметра  $A\kappa /2\pi$. Ориентация осей анизотропии повернута на 90$^o$ по сравнению со случаем изображенном на рисунке~\ref{fig_kappa_amp0}
} 
\end{figure}

Следует отметить, что основным параметром, определяющим отражение (по крайней мере, при достаточно сильной модуляции границы) является отношение поперечного (амплитуды модуляции $A$) и продольного ($L_x = 2\pi /\kappa$) масштабов модуляции, то есть параметр $g = A\kappa /2\pi$. 
На рисунке~\ref{fig_kappa_amp1} приведена зависимость $R (g )$, где красными кружками (черными звездочками) отмечены данные для пилообразного, $P=-1.2$ (косинусоидального, $P = 0.5$) профиля, и малая степень дисперсии коэффициента отражения при $g > 3.5$ указывает на то, что одинаковый эффект достигается как при сильной (относительно длинноволновой), так и слабой (но мелкомасштабной) модуляции. 

\section{Субволновая собственная мода}
\begin{figure}[htb]\centering
\includegraphics[width = 0.7\columnwidth]{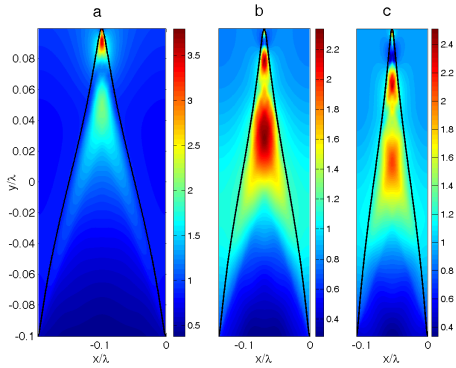}
\caption{\label{fig_mode} (color online) 
Распределение модуля магнитного поля $h(x,y)$ (показан один период структры) при нормальном падении плоской волны $h^{inc} = exp(-ik_0y)$ на периодически модулированную границу гиперболического метаматериала. Амплитуда модуляции $A=0.2\lambda$,  параметр $P$, контролирующий форму модуляции (см. формулу (\ref{eq_curve})) $P = -1.2$, период структуры в $x$ направлении $\lambda /\kappa$, где $\kappa /k_0 =  $  5, 7, и 9  соответственно для панелей (a), (b) и (c).   Ориентация осей анизотропии повернута на 90$^o$ по сравнению со случаем изображенном на рисунке~\ref{fig_fields}. 
Коэффициент отражения для этих вариантов 1\%, 4.3\% и 0.9\%  соответственно.  
} 
\end{figure}

Интересный эффект наблюдается для такой ориентации оптических осей ГМ при рассеянии падающего излучения на пилообразом профиле. 
При достаточно узких зубцах ``пилы'' поле локализуется вблизи ее вершины. Характерные распределения поля  приведены на рисунке~\ref{fig_mode}.  
Эти локализованные распределения имеют субволновой характерный масштаб, и анализ показывает, что теоретически масштаб локализации может быть сколь угодно малым. 

Действительно, в этом случае задача имеет приближенное аналитическое решение. 
Для анализа перейдем в полярную систему координат ($r$, $\phi$), с центром в вершине клина (зуба пилы) и будем считать, что диэлектрический тензор также имеет полярную симметрию, так что диагональные компоненты $\eps_{\phi\phi}= \eps_\bot <0$ и $\eps_{rr}= \eps_\| >0$. Уравнение Гельмгольца для магнитного поля в этом случае имеет вид 
\begin{align}\label{mode}
\eps_{\phi \phi}^{-1}\frac{1}{r}\frac{\pa}{\pa r} r \frac{\pa h}{\pa r} + \eps_{rr}^{-1}\frac{1}{r^2}\frac{\pa^2h}{\pa \phi^2}  + k_0^2 h =0 . 
\end{align}    
Очевидно, что субволновая локализация поля предполагает его экспоненциальное спадание в свободном пространстве от границы метаматериала, поэтому приближенно можно считать, что $h=0$ на образующих клина, и искать собственное решение в виде 
$h=H(r) \cos (\pi\phi /2\phi_0)$, где $2\phi_0$ угол раскрыва клина и угол $\phi$ отсчитывается от диагонали клина. 
Введем новую радиальную переменную $\zeta = \log r$ и запишем уравнение (\ref{mode})  с учетом выбранной угловой зависимости  
\begin{align}\label{mode1}
  \frac{d^2 H}{d\zeta^2} -  \frac{\eps_{\phi \phi}}{\eps_{rr}}\left (\frac{\pi}{2\phi_0}\right )^2H  + k_0^2\eps_{\phi \phi} e^{2\zeta} H =0 . 
\end{align}   
Первые два члена в (\ref{mode1}) дают уравнение осциллятора (напомним, что $\eps_{\phi \phi} <0$), а последнее слагаемое  мало при малых $r$ и быстро растет с увеличением расстояния от вершины клина. Критическое значение $r_{max}$, при котором последние два слагаемых становятся величинами одного порядка, можно оценить как 
$$
k_0r_{max} \approx \eps_{rr}^{-1/2}\frac{\pi}{2\phi_0}, 
$$
а при малых $r < r_{max}$ справедливо приближенное выражение 
$$
H \approx \sin \left ( \sqrt{-\eps_{\phi \phi}/\eps_{rr}}\frac{\pi}{2\phi_0} \log r \right ).
$$
В соответствии с этой формулой при уменьшении угла раскрыва клина положение максимума поля $H$ сдвигается в сторону больших $r$, что иллюстрируется на рис.~\ref{fig_mode}, где также можно заметить рост числа осцилляций поля. 
Очевидно, что число осцилляций при уменьшении $r$ стремится к бесконечности, а их характерный масштаб к нулю. Естественные ограничения на минимальный масштаб накладываются физическими причинами, т.е. поглощением в среде, шероховатостью границы метаматериала, его неоднородностью, и др. 
По-видимому, возбуждение этой собственной моды наблюдалось также в численном эксперименте \cite{cui_NanoLett12}, где также указывалось на субволновую локализацию и эффективное поглощение излучения в малом объеме.

\section{Заключение}
В качестве заключения, 
 в работе предложен метод расчета  полей, рассеянных на профилированной границе двух однородных анизотропных сред для случая, когда профиль границы является произвольной  периодической функцией координаты вдоль границы. Естественное обобщение второго тождества Грина позволило рассмотреть нетривиальный случай, когда одна из сред представляет собой гиперболический метаматериал, характеризующийся условием $\eps_{XX}\eps_{YY} <0$. В этом случае перенормировка координат превращает уравнение гиперболического типа в обычное уравнение Гельмгольца, записанное однако в комплексных пространственных переменных.  Метод работает и в поглощающей/активной среде, когда сами компоненты диэлектрического тензора являются комплексными величинами. 
 
 C помощью разработанного метода численно исследованы особенности рассеяния падающего излучения на слое гиперболического метаматериала с периодически профилированной границей, в частности, влияние формы, продольного и поперечного характерного масштаба неоднородности профиля на коэффициент отражения. Численное моделирование  показало, что достаточно мелкомасштабное профилирование границы приводит к уменьшению отражения по сравнению с плоской границей практически для всех рассмотренных форм профиля (при изменении управляющего параметра  форма профиля меняется  от почти пилообразной к почти косинусоидальной и до напоминающей меандр). Однако чем больше форма профиля отклоняется от пилообразной, тем чувствительнее оказывается коэффициент отражения к изменению амплитуды и характерного продольного масштаба модуляции.

 Найдено приближенное аналитическое выражение для  собственной моды, локализованной в гиперболическом метаматериале, который имеет пилообразный профиль границы. Теоретически сколь угодно малый масштаб локализации моды ограничивается диссипацией в среде. Определены параметры (ориентация оси анизотропии, угол раскрыва "пилы"), при которых такое (супер)субволновое решение реализуется.
 
Работа выполнена при поддержке РФФИ (грант 16-02-00556).

\section{Приложение: метод расчета полей при рассеянии плоской волны на периодически модулированной границе гиперболического метаматериала } 

Метод основан на использовании второго тождества Грина (Green's second identity) и сводит задачу рассеяния падающего излучения на границе однородной среды к вычислению поверхностных (для двумерного случая, контурных) интегралов от функции Грина и ее производной по нормали к поверхности. Похожая методика применялась в публикации \cite{Sun} для решения уравнения Гельмгольца в акустике, но в данной работе проведено обобщение, позволяющее рассматривать анизотропную среду гиперболического типа, в которой обычное (эллиптическое) уравнение Гельмгольца фактически становится  гиперболическим уравнением. 

Второе тождество Грина формулируется следующим образом. 
Если  $\psi$ и  $\phi$ обе дважды непрерывно дифференцируемы в области  $U \in R^3$, тогда 
\begin{align}\label{mapping}
  \int _{U}\left(\psi  \Delta \phi  -\phi \Delta \psi \right ) dV  
=\oint _{\partial U}  \left(\psi \frac{\partial \phi}{\partial \mathbf {n} }-\phi \frac{\partial \psi }{ \partial \mathbf {n} }\right) dS .  
\end{align}
В рассматриваемом двумерном случае лапласиан $\Delta = \partial^2/\partial x^2+ \partial^2 /\partial y^2$, интегрирование по объему заменяется на интегрирование по $x$, $y$ координатам, а граница области $\partial U$ представляет собой (параметрически заданную) линию $x_B(\tau )$, $y_B (\tau ) $ (см. (\ref{eq_curve})).

Возьмем в качестве $\phi$ рассеянное магнитное поле $h^{sct}$, а в качестве $\psi$ функцию Грина $G$ двумерной задачи, в которой точка особенности расположена на границе области $U$.

Для излучения TM поляризации уравнение  Гельмгольца в свободном пространстве имеет вид 
\begin{align}\label{simple}
\Delta h + k_0^2 h =0, 
\end{align} 
где  $h \equiv h_z$ это (единственная) $z$ компонента магнитного поля.  
Для функции Грина,  удовлетворяющей  уравнению 
\begin{align}
\Delta G + k_0^2 G  = \delta(x-x_0)\delta (y-y_0) \equiv \delta (\mathbf{r}-\mathbf{r}_0), 
\end{align}
известно аналитическое выражение  
$G(\mathbf{r} , \mathbf{r}_0) = i/4 H^1_0(k_0\rho)$, где $H^1_0$ функция Ханкеля нулевого порядка и  $\rho = \sqrt{(x-x_0)^2+(y-y_0)^2}$.  
 
Нетрудно заметить, что при таком выборе $\phi$ и $\psi$ интеграл в левой части равенства (\ref{mapping}) равен нулю, а правая часть задает соотношение между значениями рассеянного магнитного поля $h^{sct}$ и его производной по нормали $\partial h^{sct} /\partial n$ в точках на границе ($x_B(\tau_i)$, $y_B(\tau_i)$), причем аналогичное соотношение можно записать для каждой $i$-той из $N$  граничных точек, используя соответствующие функции Грина, $G_{ij} \equiv G(k_0\rho_{ij})$, где $\rho_{ij} = \sqrt{[x_B(\tau_{i})-x_B(\tau_{j})]^2+[y_B(\tau_{i})-y_B(\tau_{j})]^2}$.   
 
 Таким образом, для $2N$ неизвестных (значения $h^{sct}$ и $\partial h^{sct} /\partial n$) у нас есть $N$ уравнений, полученных из соотношений (\ref{mapping}), 
 \begin{align}\label{sum}
 \frac{\partial G_{ij}}{\partial n_j} h^{sct}_j d\tau_j -G_{ij} \frac{\partial h^{sct}_j}{\partial n_j } d\tau_j=0, 
 \end{align}
 где $d\tau_j = \sqrt{dx_j^2+dy_j^2}$  и как обычно предполагается суммирование по повторяющимся индексам. 
  Недостающие $N$ уравнений находятся из условий непрерывности тангенциальных компонент электрического и магнитного полей на границе $x_B(\tau_i)$, $y_B(\tau_i)$. 
 В простейшем случае, когда падающее излучение рассеивается на идеально отражающем металле, эти условия сводятся к равенству нулю нормальной производной полного (рассеянное плюс падающее) магнитного поля, $\partial h^{sct} /\partial n = - \partial h^{inc} /\partial n$. 
 
\begin{figure}[htb]\centering
 \includegraphics[width = 0.7\columnwidth]{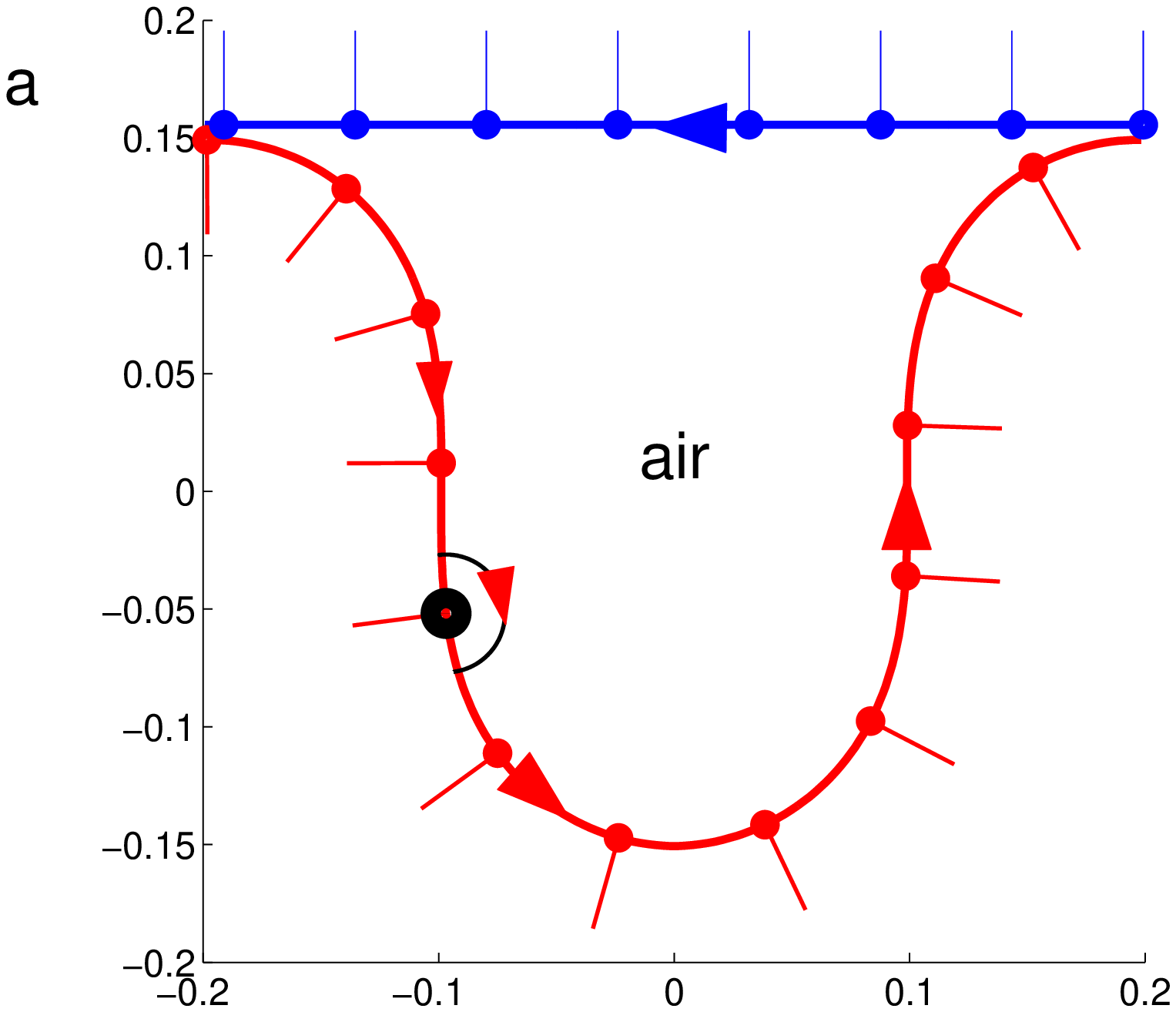}
 \includegraphics[width = 0.7\columnwidth]{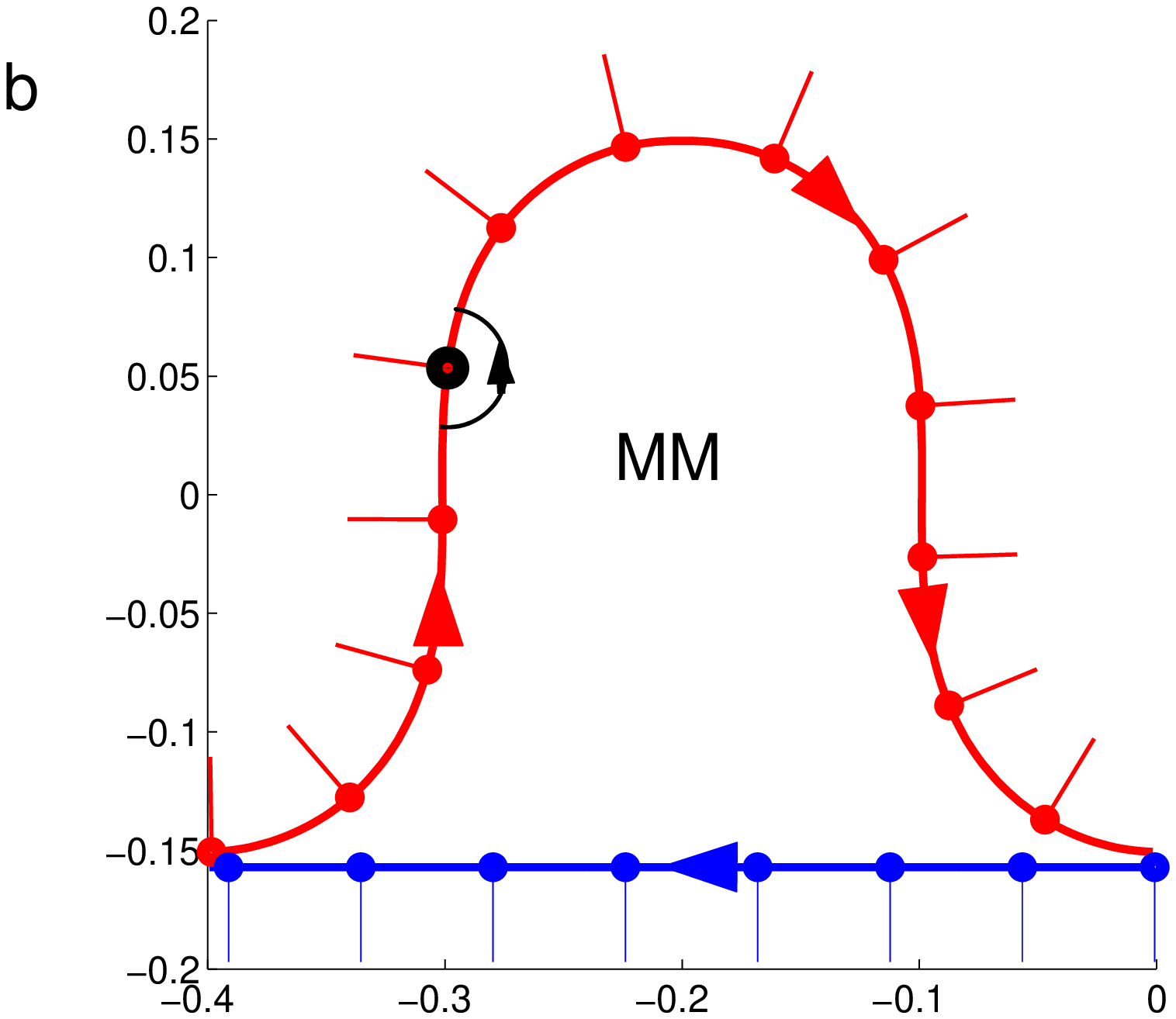}
\caption{\label{geom} (color online) 
Геометрия задачи: (a) замкнутый контур, используемый для расчета рассеянных полей в свободном пространстве, показано направление нормалей, стрелки указывают направление обхода контура. На синей части контура применяются условия радиационного излучения (см. текст); (b) то же для области метаматериала. 
} 
\end{figure}

 Если граница  $x_B(\tau )$, $y_B(\tau  )$ разделяет свободное пространство и диэлектрик с проницаемостью $\eps_D$, то в области диэлектрика волновое число $k = \sqrt{\eps_D}k_0$, и  функция Грина перенормируется как $\tilde G = G(k\rho /k_0)$.  Прошедшее магнитное поле $h^{tr}$ и его нормальная производная на границе удовлетворяет уравнению ~(\ref{mapping}) с этой новой функцией Грина $\tilde G$, а условия непрерывности полей записываются как  $h^{tr}=h^{sct}+h^{inc}$ and $\partial h^{tr}/\partial n=-\eps_D(\partial h^{sct}/\partial n+\partial h^{inc}/\partial n)$
 \footnote{
 Знак `` - '' здесь учитывает противоположное направление векторов нормали для области свободного пространства и области диэлектрика, см. рис.\ref{geom}. 
 } в каждой из  $N$ граничных точек. Таким образом, мы имеем  $4N$ уравнений для $4N$ неизвестных, что единственным образом определяет решение. 
 
 Следует отметить, что при вычислении коэффициентов матричного уравнения (\ref{sum}) (и аналогичного уравнения для поля и его нормальной производной в области диэлектрика) возникает проблема сингулярности функции Грина $G_{ii}$ и ее производной $\partial G_{ii}/\partial n_i$. 
  Очевидно, что требуется более точное вычисление диагональных коэффициентов в уравнении (\ref{sum}). Правильный результат дает   сдвиг контура интегрирования таким образом, чтобы точка особенности оставалась вне области $U$, как это иллюстрируется на рис.\ref{geom}. Соответственно,   интеграл вычисляется вдоль верхней полуокружности для полей в области свободного пространства и вдоль нижней полуокружности для полей в диэлектрике (см.~рис.~\ref{geom}).

Задача усложняется в случае, когда рассеяние происходит на границе среды с тензорной проницаемостью $\hat \eps$, например, метаматериала.  
Уравнение Гельмгольца для магнитного поля в такой среде имеет вид  
\begin{align}\label{tensor}
\frac{1}{\eps_{YY}}\frac{\partial^2 h^{tr} }{\partial X^2} + \frac{1}{\eps_{XX}}\frac{\partial^2 h^{tr} }{\partial Y^2}  + k_0^2 h^{tr} =0, 
\end{align} 
 где $X$ и $Y$ главные оси, в которых диэлектрический тензор диагонален (в общем случае оси $X,Y$ повернуты на некоторый угол $\theta$ относительно осей $x,y$, определяющих ориентацию границы: $X=x \cos \theta + y \sin \theta$, $Y=y \cos \theta - x \sin \theta$), и $\eps_{XX} = \eps_\|$, $\eps_{YY} = \eps_\bot$.

 Нетрудно заметить, что заменой 
 \begin{align}\label{change}
  \xi = \sqrt{\eps_{YY}}X ,  \  \eta = \sqrt{\eps_{XX}}Y 
 \end{align}
  уравнение~(\ref{tensor}) приводится к каноническому виду~(\ref{simple}), и функция Грина в этих переменных по-прежнему выражается через функцию Ханкеля нулевого порядка, $\tilde G = i/4 H^1_0(k_0\tilde \rho)$, где  
   $\tilde\rho = \sqrt{(\xi-\xi_0)^2+(\eta-\eta_0)^2} = \sqrt{\eps_{YY}(X-X_0)^2+\eps_{XX}(Y-Y_0)^2}$.  
 Поэтому естественным для решения задачи представляется следующий алгоритм: в метаматериальной области вводятся по формулам (\ref{change}) координаты $ \xi$, $  \eta$,  определяются граничные точки $\ \xi_B$, $ \eta_B$, соответствующие реальным граничным точкам $x_B $, $y_B $ в свободном пространстве, и вектора нормали $ \mathbf{\tilde n}$ ($\tilde n_\xi = d\eta /d\tilde\tau$, $\tilde n_\eta = -d\xi /d\tilde\tau$, где $d\tilde\tau = \sqrt{d  \xi^2 + d  \eta^2}$) в этом нормированном пространстве. Интегрирование уравнений (\ref{mapping}) в  $  \xi$, $  \eta$ пространстве дает, как и раньше,  $N$ уравнений, связывающих $h^{tr}$ and $\partial h^{tr}/\partial \tilde n = \tilde n_\xi\partial h^{tr}/\partial \xi  + \tilde n_\eta\partial h^{tr}/\partial\eta $. Заключительным шагом является  использование условий непрерывности магнитного и тангенциального электрического полей на границе между областями. 
 
 Нас интересует случай, когда диэлектрическая среда является гиперболическим метаматериалом, в котором продольная и поперечная компоненты диэлектрического тензора имеют разные знаки (для определенности будем считать, что $\eps_{XX}>0$, $\eps_{YY} <0$). 
  При этих условиях нормированные координаты $ \xi$ оказываются мнимыми, понятие вектора нормали,  как вектора единичной длины,  теряет смысл. Что более важно, эффективное расстояние до источника в функции Грина $\tilde \rho$ стремится к нулю в направлении резонансного конуса (при $(X-X_0)/(Y-Y_0) = \sqrt{-\eps_{XX}/\eps_{YY}}$), а поле в этом направлении бесконечно велико.   
  Ситуацию, как всегда, спасает  учет  затухания в метаматериале, однако при этом комплексными становятся как $ \xi$, так и $ \eta$, и интегрирование уравнения (\ref{mapping}) приходится проводить в комплексном пространстве. 
 
Однако, можно  свести вычисление коэффициентов матричного уравнения (\ref{sum})  к  интегрированию в обычном пространстве. Действительно, учтем, что 
\begin{align*}
 \frac{\partial \tilde G_{ij}}{\partial \tilde n_j}  d\tilde\tau_j =  
 \frac{\tilde G'_{ij} k_0 }{\tilde \rho_{ij}} [ (\xi_j-\xi_i)d\eta_j - (\eta_j-\eta_i)d\xi_j]  = \\  
 \sqrt{\eps_{XX}\eps_{YY}}\frac{\tilde G'_{ij} k_0 }{\tilde \rho_{ij}} [ (x_j-x_i)dy_j - (y_j-y_i)dx_j] = \\  
 \sqrt{\eps_\|\eps_\bot}\frac{\tilde G'_{ij} k_0 }{\tilde \rho_{ij}} (\mathbf{r}_j-\mathbf{r}_i)\mathbf{n}_j d\tau_j,  
\end{align*}
где $G'$ означает производную от функции Грина по аргументу. 
Аналогично можно записать выражение для  величины 
\begin{align*}
\frac{\partial h^{tr}_j}{\partial \tilde n_j } d\tilde \tau_j =  
 \frac{\partial h^{tr}_j}{\partial\xi } d\eta_j - \frac{\partial h^{tr}_j}{\partial\eta } d\xi_j =  \\
 \sqrt{\eps_\|\eps_\bot} \left (\frac{\partial h^{tr}_j}{\partial X } d Y_j \frac{1}{\eps_{YY}} 
 - \frac{\partial h^{tr}_j}{\partial Y } dX_j \frac{1}{\eps_{XX}}\right ) =   \\ 
\sqrt{\eps_{XX}\eps_{YY}}ik_0\left (E^{tr}_X dX_j + E^{tr}_Y dY_j\right ) =   
 \sqrt{\eps_\|\eps_\bot} ik_0 E^{tr}_\tau d\tau_j , 
 \end{align*}
 где $E^{tr}_\tau$ тангенциальная компонента электрического поля в прошедшей волне в граничных точках.

  Таким образом, уравнение (\ref{sum}) модифицируется для случая среды с тензорной диэлектрической проницаемостью следующим образом 
\begin{align}\label{meta}
 \sum_j \left [\frac{\tilde G'_{ij} k_0 }{\tilde \rho_{ij}} (\mathbf{r}_j-\mathbf{r}_i)\mathbf{n}_j h^{tr}_j d\tau_j - ik_0\tilde G_{ij} (E^{tr}_\tau)_j d\tau_j \right ]=0, 
 \end{align}  
 и аналогичное соотношение записывается для рассеянного поля в свободном пространстве 
  \begin{align}\label{free}
 \sum_j \left [\frac{  G'_{ij} k_0 }{  \rho_{ij}} (\mathbf{r}_j-\mathbf{r}_i)\mathbf{n}_j h^{sct}_jd\tau_j - ik_0  G_{ij} (E^{sct}_\tau)_j d\tau_j \right ] =0. 
 \end{align}
 Условие непрерывности тангенциальных электрического и магнитного полей приводит к равенствам 
 \begin{align}\label{cont}
 h^{tr}_j=h^{sct}_j+ h^{inc}_j; \qquad (E^{tr}_\tau)_j=(E^{sct}_\tau)_j+(E^{inc}_\tau)_j.
 \end{align}

 Описанная выше постановка задачи непосредственно применима для расчета рассеяния плоской волны на диэлектрическом/метаматериальном объекте конечных размеров. В этом случае граница двух сред является замкнутым контуром $\partial  U$ в формуле (\ref{mapping}) для метаматериала, а бесконечно удаленная граница свободного пространства не влияет на процесс рассеяния. 
Если же мы рассматриваем рассеяние падающего излучения на периодическом профиле и хотим ограничиться численными расчетами полей в пределах элементарной ячейки, имея в виду применение в дальнейшем теоремы Блоха, то замкнутый контур $\partial U$ в формуле (\ref{mapping}), ограничивающий область свободного пространства,  необходимо содержит участки, которые не граничат с областью метаматериала и для которых поэтому неприменимо граничное условие (\ref{cont}). Соответствующие участки  имеются также на замкнутом контуре, ограничивающем  область метаматериала (см. рис.~\ref{geom}). 
 На этих участках границы следует использовать условия радиационного излучения (условия излучения Зоммерфельда), которые связывают между собой тангенциальные компоненты электрического и магнитного полей (для свободного пространства, магнитное поле и его нормальную производную). Учет этой связи позволяет полностью рассчитать задачу и найти распределение тангенциальных компонент полей на границе $\partial U$ и тем самым коэффициент отражения. 
 
 Если же кроме коэффициента отражения нас интересует распределение полей внутри этого замкнутого контура, то для этого необходимо  найти амплитуды эффективных источников $V_i$, которые мы должны расположить лишь на части контура $\partial U$, разделяющей свободное пространство и метаматериал и которые должны создавать на этой части контура рассчитанное выше распределение полей. 
 Математически задача сводится к решению матричного уравнения 
 \begin{align*}
 V_i G_{ij} = h^{sct}_j, 
 \end{align*}
 и аналогично 
 \begin{align*}
 \tilde V_i \tilde G_{ij} = h^{tr}_j, 
 \end{align*}
 где $\mathbf{r}_i$, $\mathbf{r}_j$ принадлежат граничной части контура $\partial U$. 
 Соответственно поля в части свободного пространства и метаматериала, ограниченных контурами $\partial U$ и $\partial \tilde U$
 вычисляются как 
 $$
 h (x,y) = V_i G(k_0 \sqrt{(x-x_i)^2 + (y-y_i)^2}) + h^{inc}(x,y) ,
 $$
 $$
 \tilde h (X,Y) = \tilde V_i   G(k_0 \sqrt{\eps_{YY}(X-X_i)^2 + \eps_{XX}(Y-Y_i)^2}),
 $$
 где $h^{inc}(x,y)$ магнитное поле в падающей плоской волне.

\end{document}